\documentclass[sigconf]{acmart}

\usepackage{subcaption,booktabs} 
\usepackage{xcolor}
\usepackage[table]{xcolor}
\usepackage[linesnumbered,ruled]{algorithm2e}
\usepackage{soul}
\usepackage{hyperref}
\usepackage[
    type={CC},
    modifier={by-nc-sa},
    version={3.0},
]{doclicense}

\definecolor{light-gray}{gray}{0.95} 
\definecolor{med-gray}{gray}{0.8}

\setcopyright{none}

\acmConference[RecSys'18]{ACM Recommender Systems}{REVEAL Workshop}{Vancouver, Canada}
\acmYear{2018}
\copyrightyear{2018}

\begin{document}
\title[Offline metrics and user behaviour for combining recommendations]{Using offline metrics and user behavior analysis to combine multiple systems for music recommendation}


\author{Andres Ferraro}
\affiliation{%
  \institution{Music Technology Group - Universitat Pompeu Fabra}
  \streetaddress{Roc Boronat 138}
  \city{Barcelona}
  \state{Spain}
  \postcode{08018}
}
\email{andres.ferraro@upf.edu}

\author{Dmitry Bogdanov}
\affiliation{%
  \institution{Music Technology Group - Universitat Pompeu Fabra}
  \streetaddress{Roc Boronat 138}
  \city{Barcelona}
  \state{Spain}
  \postcode{08018}
}
\email{dmitry.bogdanov@upf.edu}

\author{Kyumin Choi}
\affiliation{%
  \institution{Kakao Corp.}
  \country{Korea}
}
\email{pi.314@kakaocorp.com}

\author{Xavier Serra}
\affiliation{%
  \institution{Music Technology Group - Universitat Pompeu Fabra}
  \streetaddress{Roc Boronat 138}
  \city{Barcelona}
  \state{Spain}
  \postcode{08018}
}
\email{xavier.serra@upf.edu}

\renewcommand{\shortauthors}{A. Ferraro et al.}

\begin{abstract}
There are many offline metrics that can be used as a reference for evaluation and optimization of the performance of recommender systems. Hybrid recommendation approaches are commonly used to improve some of those metrics by combining different systems. In this work we focus on music recommendation and propose a new way to improve recommendations, with respect to a desired metric of choice,
by combining multiple systems for each user individually based on their expected performance. 
Essentially, our approach consists in predicting an expected error that each system will produce for each user based on their previous activity.
To this end, we propose to train regression models for different metrics predicting the performance of each system based on a number of features characterizing previous user behavior in the system. We then use different fusion strategies to combine recommendations generated by each system. 
Following this approach one can optimize the final hybrid system with respect to the desired metric of choice. As a proof of concept, we conduct experiments combining two recommendation systems, a Matrix Factorization model and a popularity-based recommender. We use the data provided by \textit{Melon}, a Korean music streaming service, to train and evaluate the performance of the systems.
\end{abstract}

%
%

\ccsdesc[500]{Information systems~Recommender systems}
\ccsdesc[300]{Information systems~Information extraction}
\ccsdesc[300]{Information systems~Music retrieval}

\keywords{music recommender systems, collaborative filtering, estimation fusion, rank fusion, user modeling, offline evaluation}

\maketitle

\section{Introduction}
Music recommendation is a popular research topic within the music information retrieval (MIR) community. Collaborative filtering (CF) is one of the common solutions to this problem~\cite{celma_music_2010}. 
Many existing CF approaches to music recommendation are neighborhood-based, computing similarity between users or items~\cite{sarwar_item-based_2001,slaney_similarity_2007,celma_music_2010,aiolli_efficient_2013}. Alternatively, there were proposed model-based algorithms, in particular latent factor models based on matrix factorization techniques~\cite{koren2009matrix,koren2015advances}. In such models, users and items are simultaneously represented as feature vectors in a latent feature space learned from the user-item preference data. These methods are among the most used in recommender systems~\cite{koren2015advances} and they have been applied to music domain with a mixed success~\cite{pacula_matrix_nodate,mcfee_million_2012,volkovs_effective_2015,johnson_logistic_nodate}. Existing research lacks further evaluations of such approaches for music recommendation, comparing the performance with multiple metrics at the same time, especially in the case of using implicit user preference information such as listening statistics for artists or songs. 
%
%
 
The intrinsic problem of using implicit preference data is that, in contrast to explicit user ratings, it does not necessarily indicate relevance of consumed items and it lacks negative feedback. As a possible solution there have been proposed matrix factorization approaches specifically designed for implicit feedback, treating it as indication of preference associated with varying confidence levels~\cite{hu_collaborative_2008,rendle2009bpr}. Nevertheless, there is a lack of music studies validating and improving such approaches for the case of music recommendation. Instead, some MIR studies still use matrix factorization algorithms designed for explicit ratings data 
with implicit listening behavior information (song or artist playcounts) as an input~\cite{celma_music_2010,vigliensoni_automatic_2016,schedl_distance-_2017}. 

Regarding evaluation methodologies, many studies typically consider only a limited number of performance metrics, for example, optimizing their systems to the overall MAP~\cite{mcfee_million_2012}, or even RMSE~\cite{dror_yahoo!_2011,vigliensoni_automatic_2016,schedl_distance-_2017} (even though it is already studied that there is no direct equivalence between error metrics and precision metrics~\cite{Bellogin:2011:PER:2043932.2043996}). Thus, there is a lack of music studies with a larger variety of metrics and per-user analysis of performance of different systems in offline evaluations. 
Also, we can assume that various recommender systems may performs better or worse for different users and therefore it may be possible to select the best performing system for each user individually.

In this study we focus on addressing some of the above-mentioned research possibilities. We propose a new way to optimize the final recommendations by combining multiple systems based on predicting the error that will present each system for each user. To predict the error we train regression models for a specific metric based on previous activity of the users. 

As a proof of concept, we conduct a systematic evaluation using a variety of performance metrics on the example of two systems. One system is based on matrix factorization algorithms for collaborative filtering: the SVD algorithms for implicit data~\cite{hu_collaborative_2008}; the other system is based on popularity of the tracks. In our experiments we use data provided by \textit{Melon}, a Korean music streaming service with more than 3 million active users and a remarkably large popularity bias which makes it specially attractive for research in the context of the considered systems.



\section{Data}
We collected the listening events during a 30 days period between mid-February and mid-March of 2018 for 200,000 randomly selected \textit{Melon} users, summing up to a total of 211 million listening events. Each event has an associated user, song, and listening timestamp information. Using the metadata provided by \textit{Melon}, for each song we collected artist, album, release year, duration, as well as tags from three different taxonomies covering genre, style, and ``class'' 
referring to an alternative genre taxonomy. 
The percentage of females and males among the selected users is 52\% and 48\% respectively. Their age range varied with 27\% being between 20 and 25 years, 24\% between 26 and 30, and 23\% between 31 to 40.

One of the biggest challenges that this data offers is a strong popularity bias.
Melon offers to its users a global top-100 chart of most listened songs updated multiple times a day. 
Approximately half of gathered listening events (52\%) correspond to 111 artists, all of which occur in these charts with at least one song.

Due to high sparsity of user-song interactions we decided to focus on artist-level recommendation in our study. We extracted a user-artist playcount matrix from the listening events relying on the artist metadata of each song. The resulting data contains 23 million user-artist interactions covering approximately 126,000 different artists with a playcount matrix sparsity of 0.09\%. In this matrix the mean playcount is 8.88 with a standard deviation of 93.05. We further reduced data sparsity removing artists with less then 30 listeners. The remaining data contains $\approx$24,000 artists with the 0.48\% sparsity of the final playcount matrix. 

\begin{figure}
 \centerline{
 \includegraphics[width=\columnwidth]{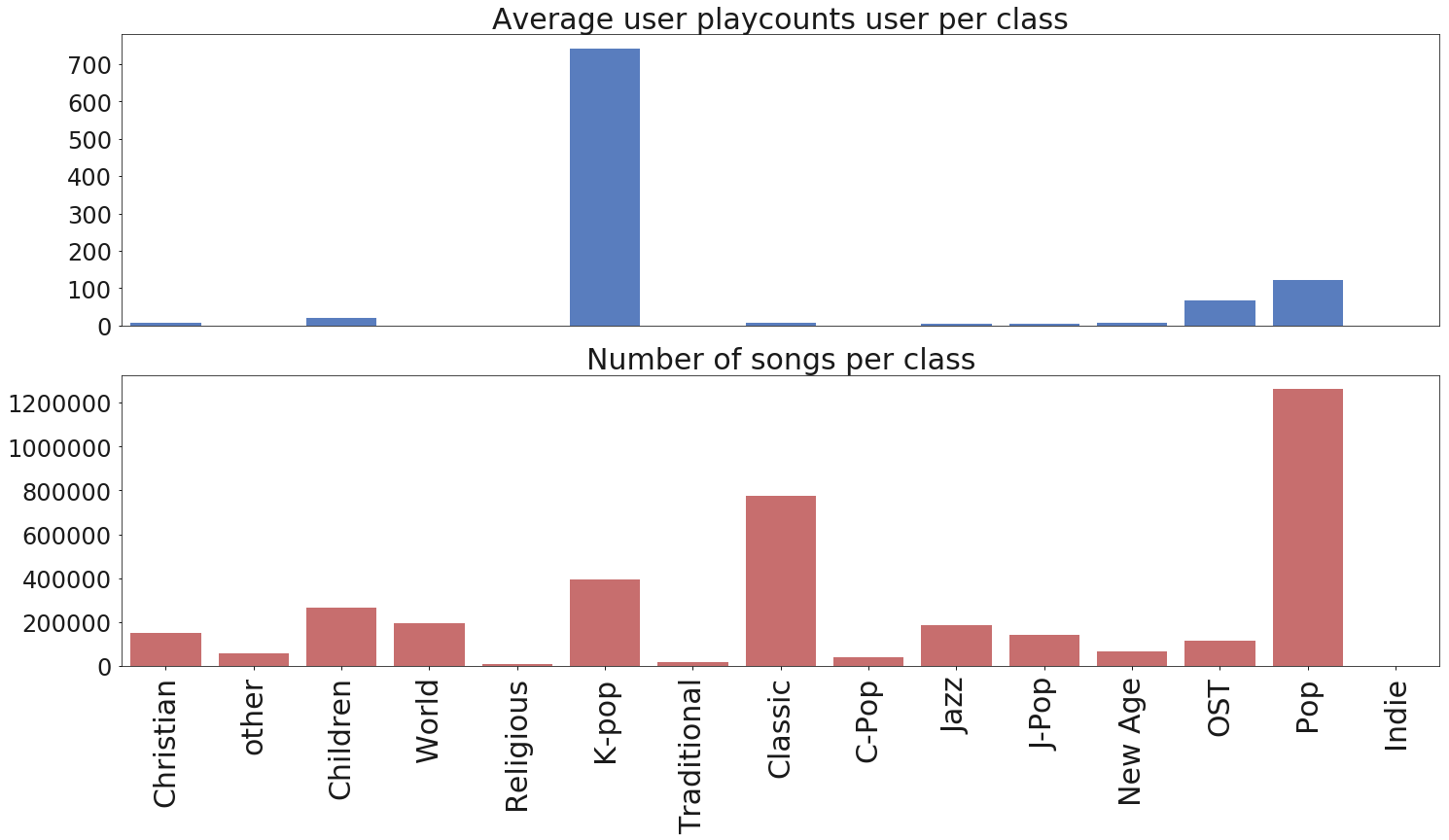}}
 \caption{Average songs listened per class (top) and number of songs per class (bottom).}
 \label{fig:class}
\end{figure}

Figure~\ref{fig:class} shows the distribution of popularity of different music classes on Melon in terms of average 
user playcounts and the number of available songs per class.
It is clear to see that the users consume mostly K-pop music. ``Pop'' is the predominant class in the dataset also being the second most listened class. 
Genre metadata can be used to improve recommendations for new users and artists, but it may be infeasible in our case given a very strong bias towards K-pop.

\subsection{Data split for experiments}\label{sec:data_split}
Similarly to the Million Song Dataset challenge~\cite{mcfee_million_2012}, we use a random 10\% subset of 20,000 users for evaluation of all systems. Their playcounts data was randomly split into an 85\% visible part (referred as the \textit{Users-Test-Visible} set), available for training systems and a 15\% hidden part for testing (\textit{Users-Test-Hidden}). For the rest of users, their entire playcount data is available for training (\textit{Users-Train}).
Hybrid systems that we propose in Section~\ref{sec:hybrid_systems} require some part of training data for estimation of expected performance of each component system, that is, their own test set. For those systems, another 20,000 users were randomly selected from the \textit{Users-Train} set. Their entire playcounts data is denoted as \textit{Users-Reg}, and it was randomly split into an 85\% train \textit{Users-Reg-Train} and 15\% test parts \textit{Users-Reg-Test}. 


\section{Evaluation of baselines systems}
\subsection{Baseline systems}
Motivated by a high popularity bias in our data, we evaluate a common collaborative filtering systems based on matrix factorization and compare the performance to popularity-based recommendations. For our experiments we selected a SVD matrix factorization algorithm commonly used in music as well as in other domains. Our baseline systems for evaluation are:
\begin{itemize}

\item \textbf{SVD-I}: A faster modification of the SVD algorithm with alternative least squares optimization specifically suited for implicit feedback data. Proposed by Hu et al.~\cite{hu_collaborative_2008} 
this algorithm is known as ``implicit SVD'' as it can be used in systems working with implicit data such as user listening behavior~\cite{zhang_auralist:_2012}. For this algorithm we used implementation from Implicit python library\footnote{\url{http://implicit.readthedocs.io}} with the following parameters: 20 factors, 0.1 regularization factor, 50 epochs, and BM25 weighting~\cite{robertson2009probabilistic} for user playcounts, as suggested by the library's documentation for the case of listening behavior data. 


\item \textbf{POP}: a popularity baseline that returns a ranked list of artists in order of decreasing overall popularity across all users.

\item \textbf{RANDOM}: a random baseline that returns a list of artists in random order.
\end{itemize}

\subsection{Evaluation metrics}
A variety of metrics can be used to measure performance of recommender systems and their choice should depend on a particular task at hand~\cite{herlocker_evaluating_2004,shani2011evaluating}. In our study we consider a scenario of retrieval of ranked lists of artist recommendations. As we work only with implicit playcount information, we need to define relevance in terms of artist playcount values. A common assumption is that higher values are associated with a higher likelihood of preference. 
We consider two cases:
\begin{itemize}
\item REL1: an artist is considered relevant for a user if it has at least one user playcount. This definition is often used in existing studies~\cite{mcfee_million_2012,Ribeiro:2014:MPA:2699158.2629350}.
\item REL10: an artist is considered as relevant if it has at least 10 playcounts.
\end{itemize}

We employ the following ranking metrics~\cite{baeza1999modern} using these relevance definitions:
\begin{itemize}
\item Precision at 10:  given a list of top 10 recommended artists, precision is defined as a fraction of recommended and relevant artists. We will refer to this metric as \textbf{$P_1@10$} and \textbf{$P_{10}@10$} for the case of REL1 and REL10 respectively. 

\item Recall at 10 and 500:  recall is defined as a fraction of a total number of relevant artists being retrieved within top 10 and 500 recommended artists respectively. Similarly, we will refer to these metrics as \textbf{$R_1@10$} and \textbf{$R_{10}@10$}, and \textbf{$R_1@500$} and \textbf{$R_{10}@500$}.

\item Mean average precision at 500~\cite{mcfee_million_2012}: given a list of top 500 recommended artists, average precision is defined as the average of precision values at all ranks where relevant artists are found. Mean average precision is an average of these values across users. Again, we consider both \textbf{$MAP_1@500$} and \textbf{$MAP_{10}@500$} depending on the definition of relevance.

\item Average percentile-rank~\cite{hu_collaborative_2008}: given a complete ranked list of artists, this metric represents an average percentile position of relevant artists in the list. Lower values are desirable as they indicate ranking relevant artist closer to the top of the recommendation lists. We consider both \textbf{$Rank_1$} and \textbf{$Rank_{10}$}. 

\item Normalized discounted cumulative gain~\cite{shani2011evaluating}: a measure of correlation between the artist ranking generated by a recommender system and the ground-truth ranking in order of decreasing user playcounts. Truncated at top 10 and 500 artist, we will refer to this measure as \textbf{$nDGG@10$} and \textbf{$nDCG@500$}, respectively. For this metric there is no need to define a threshold of playcounts, because only considers the order between the elements recommended in comparison with the order of the elements in the ground truth.
\end{itemize}

In addition we consider two metrics to assess diversity of recommendations:
\begin{itemize}
\item Diversity ($DIV$): Percentage of different artists among all top-500 recommendation lists for all users.

\item Repetition ($REP$): Average number of users reached per artist recommendation among all artists in top-500 recommendation lists.
\end{itemize}

\subsection{Training and testing data}
Following the data split described in Section~\ref{sec:data_split} we used \textit{User-Train} and \textit{User-Test-Visible} data subsets for training SVD-I baseline and for computing global popularity for POP. All baselines were evaluated on \textit{User-Test-Hidden} data.

\subsection{Results}
The evaluation results are presented in Table~\ref{table:results_baselines}. We observe a consistently better overall performance of SVD-I compared to POP except for the $Rank_{10}$ metric. As expected, SVD-I produces much more diverse artist recommendations according to $DIV$ and $REP$ metrics. Still, per-user  analysis of results suggest that using POP recommendations results in better relevance metrics for 4.6\%-33.67\% of the users depending on the metric. This observation reassures the existence of a high popularity bias in our dataset. It demonstrates that when optimizing relevance with no regard to diversity one may end up using a popularity-based system for a large part of the users of a system.

\section{Hybrid systems}\label{sec:hybrid_systems}
In general we can assume that various recommender systems may performs better or worse for different users and therefore it may be possible to select the best performing system for each user individually. This decision can be done relying on an estimation of expected performance of each system for each user on a part of user data. Alternatively, ranked recommendation lists generated by various systems can be combined using simple rank fusion techniques with a weight for each system based on the estimated performance (better systems receive greater weights).
To obtain such estimations, we decided to investigate the relationship between patterns of user listening behavior and performance metrics of recommender systems.  We hypothesize that, at least for some systems, their performance will be similar for users with similar listening patterns. Related studies have suggested that incorporating such behavioral patterns may improve CF recommendation~\cite{herrera_rocking_2010,schedl_tailoring_2015,vigliensoni_automatic_2016}. 

\subsection{User behavior features}
We characterized listening behavior of \textit{Melon} users by a number of features extracted from their listening history and studied how these features can be used to predict performance of recommender systems using linear regression models. Using the timestamp information we divided listening history of each user in sessions. For each user a session is defined as a set of songs listened sequentially with a time gap between the end of one and the start of the next song no longer than 15 minutes. Some of the user features are computed for each session and then summarized across sessions by mean and standard deviation while others are computed on the entire set of songs listened by a user. 
The session features include:
\begin{itemize}
\item Percentage of repeated artists, songs, albums, genres, style, class and year out of total number of songs listened in a session including duplicates.
\item Exploratoryness~\cite{vigliensoni_automatic_2016} computed for a session.
\item Percentage of songs listened completely and percentage skipped.
\item Percentage of the session time that a user was inactive (not listening anything).
\item Average song duration percentage actually listened in a session.
\item Percentage of changes (or jumps) between artists, albums, genres, styles and classes between consecutive tracks.
\item Average popularity percentile-rank of all artists listened in a session, this measure also considers artist listened multiple times in a session.
\item Average artist mainstreamness~\cite{vigliensoni_automatic_2016}.
\end{itemize}

In turn, the features computed for the entire history of each user include:
\begin{itemize}
\item Percentage of repeated artists, songs, albums, genres, style, class and year.
\item Distribution of the listened songs for each decade, class and genre.
\item Percentage of listened songs from their total amount for each hour/weekday.
\end{itemize}

In our study these features are treated as an external data source. They are computed for all users from \textit{User-Train} and \textit{User-Test} 
from an extended dataset.

\subsection{Hybrid systems based on regression models predicting performance of component systems}
Having observed a notable performance of POP compared to SVD-I for a part of test users, we decided to take further advantage of both baselines by combining them into a number of hybrid systems. 

In our hybrid systems, we propose to rely on linear regression models able to predict expected performance metrics for each component system (SVD-I and POP) given user behavior features. To this end, counterparts to SVD-I and POP are trained on a conjunction of \textit{User-Test} and \textit{User-Train} excluding \textit{User-Reg-Test} which is used to estimate their performance. Thereby we ensure to use the maximum amount of information for training the counterpart systems in attempt to avoid a potential bias due to data size. Once the evaluation of their performance is done for each user in the \textit{Users-Reg} set, linear regression models are trained to map behavior features of those users to the estimated values of ranking metrics. A separate regression model is created for each metric and counterpart system (24 models in total). 

Table~\ref{tab:r2values} presents the obtained R-squared values for all regression models. As we can see, $Rank_{10}$ and $Rank_1$ were the most difficult metrics to predict. No correlation was found between user features and $Rank_{10}$ for both systems. For other metrics the models were able to explain from 23\% to 59\% of variability. The best results were obtained predicting $MAP_1@500$, $nDCG@500$, $R_1@10$, $P_1@10$  and $P_{10}@10$ ($\geq$ 40\% of explained variability).

\begin{table}[!ht]
\footnotesize
\setlength{\tabcolsep}{3pt}
\begin{center}
\begin{tabular}{lcc}
\toprule
  Regression model & SVD-I  & POP \\
\midrule
$P_{1}@10$ & \textbf{0.483} & 0.521 \\
$P_{10}@10$ & 0.441 & 0.456 \\
$MAP_{1}@500$ & 0.456 & 0.535 \\
$MAP_{10}@500$ & 0.271 & 0.417 \\ 
$R_{1}@10$ & 0.413 & 0.418 \\
$R_{10}@10$ & 0.234 & 0.326 \\
$R_{1}@500$ & 0.367 & 0.512 \\
$R_{10}@500$ & 0.098 & 0.258 \\
$Rank_{1}$ & 0.077 & 0.308 \\
$Rank_{10}$ & 0.047 & 0.040 \\
$nDCG@500$ & 0.395 & \textbf{0.589} \\
$nDCG@10$ & 0.339 & 0.517 \\
\bottomrule
 \end{tabular}
\end{center}
 \caption{R-squared value for regression models predicting system performance using user behavior features.}
 \label{tab:r2values}
\end{table}

Based on the pre-trained regression models, the hybrid systems proceed with the following steps for generating recommendations:
\begin{enumerate}
\item For each user, predict expected performance of each component system (SVD-I and POP) by regression models given the user's behavior features.
\item Based on the predicted metric values we considered two simple approaches: 
\begin{itemize}
\item Select the best system for this user according to the considered metric (\textbf{SELECT}) and generate a ranked recommendation list.
\item Apply a weighted ranking fusion~\cite{zhang_fusion-based_2010} combining the ranked recommendation lists from both systems. The weights are defined either individually for each user (\textbf{FUSE}) or globally being averaged across all test users (\textbf{FUSE-Avg}). 
For each system ($p$ for POP, $s$ for SVD-I), depending on the metric that we want to optimize ($m_{s}$), we calculate the values of the weights as:\footnote{In the case of $Rank_1$ and $Rank_{10}$ smaller values signify better performance, therefore $1-m_s$ and $1-m_p$ were used instead of $m_s$ and $m_p$.}
   
\[   
w_{s} = \frac{m_{s}}{m_{s}+m_{p}}, w_{p} = \frac{m_{p}}{m_{s}+m_{p}}
\]

Then for each user ($u$) we use these weights to combine the rankings of each system ($r_{u,s}$ or $r_{u,p}$):

\[
r_{u,h} = \frac{r_{u,p} * w_{p} + r_{u,s} * w_{s}}{2}
\]

\end{itemize}
\end{enumerate}
In total, we consider 36 systems combining three methods for merging systems (SELECT, FUSE, FUSE-Avg) and optimization with respect to 12 ranking metrics.  

\subsection{Evaluation results}
Table~\ref{table:results_hybrid} presents evaluation results on the \textit{Users-Test-Hidden} set. Overall, the best performance values are always achieved by hybrid methods, except for $R_1@500$ and $R_{10}@500$ for which no hybrid system could outperform the SVD-I baseline. Similarly, it was impossible to surpass the baselines in terms of $Rank_{10}$, except for the SELECT systems optimized with respect to Rank metrics.
 
Remarkably, FUSE-nDCG@500 achieved the majority of best metric values: it improves all ranking metrics to their best except for Ranks and recalls at 500, and it also provides a decent improvement in diversity compared to the baselines. We can conclude that it may be beneficial to prioritize the $nDCG@500$ metric when optimizing systems. Similarly, FUSE-MAP1@500 also performed very well: it achieved the best performance in terms of precision, MAP and $R_1@10$ measures, and improved on the rest of the metrics except for recall and $Rank_{10}$.


The results also suggest that ranking fusion systems have better performance than SELECT systems. There is not much difference in ranking metrics for FUSE personalized for each user individually and FUSE-Avg using the same weights for all users. However, these methods differ in diversity and repetition: personalized weights allow much greater diversity surpassing the baselines.
This is a remarkable results since we combine collaborative filtering with very popular recommendations, exploiting the popularity bias, but also improving the diversity at the same time.



In Table~\ref{sysusers} we show the percentage of users selected for each hybrid system from each baseline (SVD-I and POP). We can see that in some cases the percentage is similar for both systems or slighly higher in the case of SELECT-Rank1 for POP, but in most cases there is a clear superiority of SVD-I. Comparing these results with the values of the metrics in Table~\ref{table:results_baselines}  it is clear that the POP has better performance in Rank1 and accordingly most users are selected to use this system. On the opposite, for $R_1@500$ and $R_{10}@500$ there is apparent large superiority of SVD-I, so is not possible to improve the recommendations for this metric in any of the hybrid system.

\begin{table}[!ht]
 \footnotesize
 \setlength{\tabcolsep}{3pt}
 \begin{center}
 \begin{tabular}{lcc}
 
  \toprule
  
  System & \textbf{ \% users SVD-I}  & \textbf{ \% users POP} \\
\midrule
SELECT-P1@10  & 77.40  & 22.60 \\
SELECT-P10@10  & 78.61  & 21.39 \\
SELECT-MAP1@500  & 83.02  & 16.97 \\
SELECT-MAP10@500  & 86.14  & 13.86 \\
SELECT-R1@10  & 75.73  & 24.27 \\
SELECT-R10@10  & 78.36  & 21.63 \\
SELECT-R1@500  & \textbf{96.41}  & 3.59 \\
SELECT-R10@500  & 84.92  & 15.08 \\
SELECT-Rank1  & 68.68  & 31.32 \\
SELECT-Rank10  & 42.79  & \textbf{57.20} \\
SELECT-nDCG@500  & 92.94  & 7.06 \\
SELECT-nDCG@10  & 84.68  & 15.32 \\
\bottomrule
 \end{tabular}
\end{center}
 \caption{Number of users for each selected baseline approach in SELECT hybrid systems.}
 \label{sysusers}
\end{table}

\begin{table*}
 \footnotesize
 \setlength{\tabcolsep}{3pt}
 \begin{subtable}{\linewidth}
 \centering
  \begin{tabular}{p{3cm}cccccccccccccc}
\toprule
System & \textbf{$P_1$}  & \textbf{$P_{10}$}  & \textbf{$MAP_1$}  & \textbf{$MAP_{10}$}  & \textbf{$R_1$}  & \textbf{$R_{10}$}  & \textbf{$R_1$}  & \textbf{$R_{10}$}  & \textbf{$Rank_1$}  & \textbf{$Rank_{10}$}  & \textbf{$nDCG$}  & \textbf{$nDCG$}  & \textbf{$DIV$}  & \textbf{$REP$} \\

& \textbf{$@10$}  & \textbf{$@10$}  & \textbf{$@500$}  & \textbf{$@500$}  & \textbf{$@10$}  & \textbf{$@10$}  & \textbf{$@500$}  & \textbf{$@500$}  &   &   & \textbf{$@500$}  & \textbf{$@10$}  &  &  \\
\midrule  
RANDOM & 0.001 & 0.000 & 0.004 & 0.001 & 0.000 & 0.000 & 0.020 & 0.020 & 0.499 & 0.518 & 0.005 & 0.000 & 100 & 314 \\
POP & 0.227 & 0.099 & 0.186 & 0.162 & 0.141 & 0.263 & 0.697 & 0.819 & 0.036 & \cellcolor{med-gray}0.071 & 0.340 & 0.196 & 3.77 & 8325 \\
SVD-I &  \cellcolor{med-gray}0.268 &  \cellcolor{med-gray}0.123 &  \cellcolor{med-gray}0.233 &  \cellcolor{med-gray}0.227 &  \cellcolor{med-gray}0.175 &  \cellcolor{med-gray}0.339 &  \cellcolor{med-gray}0.793 &  \cellcolor{med-gray}0.908 &  \cellcolor{med-gray}0.027 & 0.078 &  \cellcolor{med-gray}0.425 &  \cellcolor{med-gray}0.269 & 30.45 &  1032 \\
\bottomrule
\end{tabular}
\caption{Baselines}
\label{table:results_baselines}
 \end{subtable}
 
 \vspace{0.5cm}
 \begin{subtable}{\linewidth}
 \centering
 \begin{tabular}{p{3cm}cccccccccccccc}
 
  \toprule

System & \textbf{$P_1$}  & \textbf{$P_{10}$}  & \textbf{$MAP_1$}  & \textbf{$MAP_{10}$}  & \textbf{$R_1$}  & \textbf{$R_{10}$}  & \textbf{$R_1$}  & \textbf{$R_{10}$}  & \textbf{$Rank_1$}  & \textbf{$Rank_{10}$}  & \textbf{$nDCG$}  & \textbf{$nDCG$}  & \textbf{$DIV$}  & \textbf{$REP$} \\

& \textbf{$@10$}  & \textbf{$@10$}  & \textbf{$@500$}  & \textbf{$@500$}  & \textbf{$@10$}  & \textbf{$@10$}  & \textbf{$@500$}  & \textbf{$@500$}  &   &   & \textbf{$@500$}  & \textbf{$@10$}  &  &  \\

\midrule
  
SELECT-P1@10 & \cellcolor{light-gray}0.277 & \cellcolor{light-gray}0.125 & \cellcolor{light-gray}0.236 & \cellcolor{med-gray}0.227 & \cellcolor{light-gray}0.179 & \cellcolor{light-gray}0.341 & 0.774 & 0.887 & 0.028 & 0.079 & 0.418 &\cellcolor{light-gray} 0.270 & 29.96 & 1049 \\
FUSE-P1@10 & \cellcolor{light-gray}0.315 & \cellcolor{light-gray}0.141 & \cellcolor{light-gray}0.262 & \cellcolor{light-gray}0.270 & \cellcolor{light-gray}0.196 & \cellcolor{light-gray}0.377 & 0.768 & 0.881 & \cellcolor{med-gray}0.027 & 0.089 & \cellcolor{light-gray}0.449 & 0.318 & \cellcolor{light-gray}80.42 & \cellcolor{light-gray}391 \\
FUSE-AVG-P1@10 & \cellcolor{light-gray}0.315 & \cellcolor{light-gray}0.141 & \cellcolor{light-gray}0.263 & \cellcolor{light-gray}0.271 & \cellcolor{light-gray}0.197 & \cellcolor{light-gray}0.380 & 0.779 & 0.893 & \cellcolor{light-gray}\textbf{0.023} & 0.087 & \cellcolor{light-gray}0.454 & \cellcolor{light-gray}0.320 & 11.88 & 2646\\
\midrule
SELECT-P10@10 & \cellcolor{light-gray}0.272 & \cellcolor{light-gray}0.125 & \cellcolor{light-gray}0.233 & \cellcolor{med-gray}0.227 & \cellcolor{light-gray}0.176 & \cellcolor{light-gray}0.342 & 0.775 & 0.892 & 0.028 & 0.077 & 0.419 & \cellcolor{light-gray}0.270 & 30.13 & 1043 \\
FUSE-P10@10 & \cellcolor{light-gray}0.311 & \cellcolor{light-gray}0.140 & \cellcolor{light-gray}0.258 & 0.268 & 0.192 & \cellcolor{light-gray}0.374 & 0.760 & 0.871 & 0.029 & 0.090 & \cellcolor{light-gray}0.445 & \cellcolor{light-gray}0.316 & \cellcolor{light-gray}84.96 & \cellcolor{light-gray}370 \\
FUSE-AVG-P10@10 & \cellcolor{light-gray}0.315 & \cellcolor{light-gray}0.141 & \cellcolor{light-gray}0.263 & \cellcolor{light-gray}0.271 & \cellcolor{light-gray}0.197 & \cellcolor{light-gray}0.381 & 0.780 & 0.894 & \cellcolor{light-gray}0.024 & 0.086 & \cellcolor{light-gray}0.454 & \cellcolor{light-gray}0.320 & 12.14 & 2590\\
\midrule
SELECT-MAP1  & \cellcolor{light-gray}0.277 & \cellcolor{light-gray}0.125 & \cellcolor{light-gray}0.238 & \cellcolor{light-gray}0.231 & \cellcolor{light-gray}0.179 & \cellcolor{light-gray}0.343 & 0.777 & 0.889 & 0.029 & 0.080 & 0.421 & \cellcolor{light-gray}0.273 & 29.96 & 1049 \\
FUSE-MAP1 & \cellcolor{light-gray}\textbf{0.316} & \cellcolor{light-gray}\textbf{0.142} & \cellcolor{light-gray}\textbf{0.264} & \cellcolor{light-gray}\textbf{0.272} & \cellcolor{light-gray}\textbf{0.198} & \cellcolor{light-gray}0.381 & 0.774 & 0.888 & \cellcolor{light-gray}0.026 & 0.088 & \cellcolor{light-gray}0.453 & \cellcolor{light-gray}0.321 & \cellcolor{light-gray}74.05 & \cellcolor{light-gray}424 \\
FUSE-AVG-MAP1 & \cellcolor{light-gray}0.315 & \cellcolor{light-gray}0.141 & \cellcolor{light-gray}0.263 & \cellcolor{light-gray}0.271 & \cellcolor{light-gray}0.197 & \cellcolor{light-gray}0.381 & 0.780 & 0.894 & \cellcolor{light-gray}0.024 & 0.086 & \cellcolor{light-gray}0.454 & \cellcolor{light-gray}0.320 & 12.14 & 2590 \\ 
\midrule
SELECT-MAP10  & \cellcolor{light-gray}0.274 & \cellcolor{light-gray}0.125 & \cellcolor{light-gray}0.236 & \cellcolor{light-gray}0.231 & \cellcolor{light-gray}0.178 & \cellcolor{light-gray}0.345 & 0.779 & 0.895 & 0.028 & 0.080 & 0.423 & \cellcolor{light-gray}0.275 & 30.01 & 1047\\
FUSE-MAP10 & \cellcolor{light-gray}0.310 & \cellcolor{light-gray}0.140 & \cellcolor{light-gray}0.257 & 0.267 & 0.192 & 0.373 & 0.756 & 0.868 & 0.031 & 0.093 & \cellcolor{light-gray}0.444 & \cellcolor{light-gray}0.315 & \cellcolor{light-gray}89.57 & \cellcolor{light-gray}351 \\
FUSE-AVG-MAP10 & \cellcolor{light-gray}0.315 & \cellcolor{light-gray}0.141 & \cellcolor{light-gray}0.263 & \cellcolor{light-gray}0.271 & \cellcolor{light-gray}0.197 & \cellcolor{light-gray}0.381 & 0.782 & 0.896 & \cellcolor{light-gray}0.024 & 0.086 & \cellcolor{light-gray}0.454 & \cellcolor{light-gray}0.320 & 12.67 & 2482\\
\midrule
SELECT-R1@10 & \cellcolor{light-gray}0.276 & \cellcolor{light-gray}0.124 & \cellcolor{light-gray}0.236 & \cellcolor{light-gray}0.228 & \cellcolor{light-gray}0.179 & \cellcolor{light-gray}0.341 & 0.770 & 0.885 & 0.030 & 0.080 & 0.417 & \cellcolor{light-gray}0.271 & 30.17 & 1042 \\
FUSE-R1@10 & \cellcolor{light-gray}0.312 & \cellcolor{light-gray}0.140 & \cellcolor{light-gray}0.261 & 0.269 & 0.197 & \cellcolor{light-gray}0.378 & 0.770 & 0.881 & \cellcolor{med-gray}0.027 & 0.089 & \cellcolor{light-gray}0.449 & 0.318 & \cellcolor{light-gray}81.40 & \cellcolor{light-gray}386 \\
FUSE-AVG-R1@10 & \cellcolor{light-gray}0.315 & \cellcolor{light-gray}0.141 & \cellcolor{light-gray}0.263 & \cellcolor{light-gray}0.271 & \cellcolor{light-gray}0.197 & \cellcolor{light-gray}0.381 & 0.780 & 0.894 & \cellcolor{light-gray}\textbf{0.023} & 0.087 & \cellcolor{light-gray}0.454 & \cellcolor{light-gray}0.320 & 12.05 & 2609\\
\midrule
SELECT-R10@10 & \cellcolor{light-gray}0.272 & \cellcolor{light-gray}0.125 & \cellcolor{light-gray}0.233 & \cellcolor{light-gray}0.227 & \cellcolor{light-gray}0.176 & \cellcolor{light-gray}0.343 & 0.774 & 0.891 & 0.029 & 0.078 & 0.419 & \cellcolor{light-gray}0.271 & 30.06 & 1045 \\
FUSE-R10@10 & \cellcolor{light-gray}0.311 & \cellcolor{light-gray}0.139 & \cellcolor{light-gray}0.259 & 0.268 & 0.194 & \cellcolor{light-gray}0.376 & 0.761 & 0.873 & 0.029 & 0.092 & \cellcolor{light-gray}0.446 & 0.316 & \cellcolor{light-gray}84.89 & \cellcolor{light-gray}370 \\
FUSE-AVG-R10@10 & \cellcolor{light-gray}0.315 & \cellcolor{light-gray}0.141 & \cellcolor{light-gray}0.263 & \cellcolor{light-gray}0.271 & \cellcolor{light-gray}0.197 & \cellcolor{light-gray}0.381 & 0.780 & 0.895 & \cellcolor{light-gray}0.024 & 0.086 & \cellcolor{light-gray}0.454 & \cellcolor{light-gray}0.321 & 12.28 & 2560\\
\midrule
SELECT-R10@500 & \cellcolor{light-gray}0.258 & 0.120 & 0.222 & 0.216 & 0.167 & 0.330 & 0.786 & 0.906 & 0.026 & 0.075 & 0.415 & 0.258 & 30.42 & 1033 \\
FUSE-R10@500 & \cellcolor{light-gray}0.315 & \cellcolor{light-gray}0.141 & \cellcolor{light-gray}0.262 & \cellcolor{light-gray}0.270 & \cellcolor{light-gray}0.197 & \cellcolor{light-gray}0.380 & 0.780 & 0.895 & \cellcolor{light-gray}\textbf{0.023} & 0.086 & \cellcolor{light-gray}0.453 & \cellcolor{light-gray}0.320 & 31.46 & 999 \\
FUSE-AVG-R10@500 & \cellcolor{light-gray}0.314 & \cellcolor{light-gray}0.141 & \cellcolor{light-gray}0.263 & \cellcolor{light-gray}0.270 & \cellcolor{light-gray}0.197 & \cellcolor{light-gray}0.379 & 0.778 & 0.892 & 0.024 & 0.087 & \cellcolor{light-gray}0.453 & 0.320 & 11.57 & 2718\\
\midrule
SELECT-R1@500 & \cellcolor{light-gray}0.266 & 0.122 & \cellcolor{light-gray}0.229 & 0.223 & 0.172 & 0.337 & 0.792 & 0.907 & \cellcolor{light-gray}0.026 & 0.077 & 0.422 & 0.266 & 30.45 & 1033 \\
FUSE-R1@500 & \cellcolor{light-gray}0.315 & \cellcolor{light-gray}0.141 & \cellcolor{light-gray}0.263 & \cellcolor{light-gray}0.271 & \cellcolor{light-gray}0.197 & \cellcolor{light-gray}0.381 & 0.781 & 0.895 & \cellcolor{light-gray}\textbf{0.023} & 0.086 & \cellcolor{light-gray}0.454 & \cellcolor{light-gray}0.320 & 25.42 & 1236 \\
FUSE-AVG-R1@500 & \cellcolor{light-gray}0.315 & \cellcolor{light-gray}0.141 & \cellcolor{light-gray}0.263 & \cellcolor{light-gray}0.271 & \cellcolor{light-gray}0.197 & \cellcolor{light-gray}0.379 & 0.778 & 0.892 & \cellcolor{light-gray}0.024 & 0.087 & \cellcolor{light-gray}0.453 & \cellcolor{light-gray}0.320 & 11.68 & 2691 \\
\midrule
SELECT-Rank1 & \cellcolor{light-gray}0.255 & 0.117 & 0.214 & 0.203 & 0.160 & 0.316 & 0.778 & 0.895 & 0.025 & \cellcolor{light-gray}0.070 & 0.401 & 0.244 & 30.38 & 1034 \\
FUSE-Rank1 & \cellcolor{light-gray}0.298 & \cellcolor{light-gray}0.135 & \cellcolor{light-gray}0.246 & \cellcolor{light-gray}0.252 & \cellcolor{light-gray}0.183 & 0.357 & 0.757 & 0.868 & 0.028 & 0.087 & \cellcolor{light-gray}0.433 & \cellcolor{light-gray}0.299 & 71.16 & \cellcolor{light-gray}441 \\
FUSE-AVG-Rank1 & \cellcolor{light-gray}0.315 & \cellcolor{light-gray}0.141 & \cellcolor{light-gray}0.263 & \cellcolor{light-gray}0.271 & \cellcolor{light-gray}0.197 & \cellcolor{light-gray}0.381 & 0.782 & 0.896 & \cellcolor{light-gray}0.024 & 0.086 & \cellcolor{light-gray}0.455 & \cellcolor{light-gray}0.320 & 12.70 & 2475\\
\midrule
SELECT-Rank10 & \cellcolor{light-gray}0.241 & 0.110 & 0.201 & 0.184 & 0.150 & 0.289 & 0.747 & 0.871 & 0.028 & \cellcolor{light-gray}\textbf{0.067} & 0.376 & 0.221 & 30.22 & 1040 \\
FUSE-Rank10 & \cellcolor{light-gray}0.296 & \cellcolor{light-gray}0.130 & \cellcolor{light-gray}0.247 & \cellcolor{light-gray}0.253 & \cellcolor{light-gray}0.185 & \cellcolor{light-gray}0.358 & 0.747 & 0.859 & 0.031 & 0.092 & \cellcolor{light-gray}0.429 & \cellcolor{light-gray}0.300 & \cellcolor{light-gray}89.64 & \cellcolor{light-gray}\textbf{350} \\
FUSE-AVG-Rank10 & \cellcolor{light-gray}0.314 & \cellcolor{light-gray}0.140 & \cellcolor{light-gray}0.261 & \cellcolor{light-gray}0.268 & \cellcolor{light-gray}0.197 & \cellcolor{light-gray}0.376 & 0.775 & 0.889 & \cellcolor{light-gray}0.024 & 0.085 & \cellcolor{light-gray}0.449 & \cellcolor{light-gray}0.317 & 10.87 & 2892\\
\midrule
SELECT-nDCG@500 & \cellcolor{light-gray}0.270 & \cellcolor{light-gray}0.124 & \cellcolor{light-gray}0.234 & \cellcolor{light-gray}0.230 & \cellcolor{light-gray}0.176 & \cellcolor{light-gray}0.344 & 0.788 & 0.904 & \cellcolor{med-gray}0.027 & 0.079 & \cellcolor{light-gray}0.426 & \cellcolor{light-gray}0.273 & 30.37 & 1035 \\
FUSE-nDCG@500 & \cellcolor{light-gray}\textbf{0.316} & \cellcolor{light-gray}\textbf{0.142} & \cellcolor{light-gray}\textbf{0.264} & \cellcolor{light-gray}\textbf{0.272} & \cellcolor{light-gray}\textbf{0.198} & \cellcolor{light-gray}\textbf{0.383} & 0.780 & 0.895 & \cellcolor{light-gray}0.024 & 0.087 & \cellcolor{light-gray}\textbf{0.456} & \cellcolor{light-gray}\textbf{0.322} & \cellcolor{light-gray}56.69 & \cellcolor{light-gray}554 \\
FUSE-AVG-nDCG@500 & \cellcolor{light-gray}0.315 & \cellcolor{light-gray}0.141 & \cellcolor{light-gray}0.263 & \cellcolor{light-gray}0.271 & \cellcolor{light-gray}0.197 & \cellcolor{light-gray}0.381 & 0.780 & 0.894 & \cellcolor{light-gray}0.024 & 0.086 & \cellcolor{light-gray}0.454 & \cellcolor{light-gray}0.320 & 12.14& 2588\\
\midrule
SELECT-nDCG@10 & \cellcolor{light-gray}0.274 & \cellcolor{light-gray}0.125 & \cellcolor{light-gray}0.236 & \cellcolor{light-gray}0.231 & \cellcolor{light-gray}0.178 & \cellcolor{light-gray}0.345 & 0.778 & 0.895 & 0.028 & 0.079 & \cellcolor{light-gray}0.423 & \cellcolor{light-gray}0.275 & 29.98 & 1048 \\
FUSE-nDCG@10 & \cellcolor{light-gray}0.310 & \cellcolor{light-gray}0.140 & 0.257 & \cellcolor{light-gray}0.266 & \cellcolor{light-gray}0.191 & \cellcolor{light-gray}0.373 & 0.754 & 0.866 & 0.030 & 0.093 & \cellcolor{light-gray}0.443 & \cellcolor{light-gray}0.315 & \cellcolor{light-gray}\textbf{89.76} & \cellcolor{light-gray}\textbf{350} \\
FUSE-AVG-nDCG@10 & \cellcolor{light-gray}0.315 & \cellcolor{light-gray}0.141 & \cellcolor{light-gray}0.263 & \cellcolor{light-gray}0.271 & \cellcolor{light-gray}\textbf{0.198} & \cellcolor{light-gray}0.381 & 0.781 & 0.896 & \cellcolor{light-gray}0.024 & 0.086 & \cellcolor{light-gray}0.454 & \cellcolor{light-gray}0.321 & 12.57& 2500\\
\bottomrule
 \end{tabular}
 \caption{Hybrid methods}
 \label{table:results_hybrid}
  \end{subtable}
 \caption{Evaluation results for the baselines and the proposed hybrid systems. Values highlighted in dark gray correspond to the best performance achieved by the baselines. Values in gray correspond to improvements over the baselines. Values in bold mark the best performance achieved.}
 \label{table:results}
\end{table*}

\section{Conclusions}
In this study we first evaluated two baseline recommendation systems, one based on matrix factorization and another that produces recommendations based on popularity, using a variety of offline metrics for performance. For part of the users, popularity-based recommendations outperformed collaborative filtering, and therefore we proposed to combine both approaches together. Ranking-fusion hybrid systems with personalized weights for each user were able to improve on ranking metrics by exploiting popularity bias, but also improving the diversity of recommendations. We propose to compute such weights based on prediction of the expected performance of each component system. We showed that such prediction can be done by regression models individually for each user using her/his listening behavior patterns as an input. To this end, we considered a number of user behavior features some of which were previously suggested in related studies~\cite{schedl_distance-_2017,vigliensoni_automatic_2016}.  

In contrast to many studies, our evaluations were conducted on an extended set of performance metrics. We found that optimizing our ranking-fusion hybrid systems for nDCG@500 led to the best overall results surpassing the baselines on the majority of metrics. Also, our method allows to optimize the systems according to the desired metric in most of the cases.

For the future work we propose to study alternative ranking-fusion or score-fusion methods. It is also challenging to estimate the importance of individual user behavior features and compare global features and session features for predicting system performance\footnote{In pre-analysis we identified a number of important features, such as the percentage of repeated artist which performs better than the others for predicting the performance. We also saw a clear improvement by calculating this feature by sessions in comparison with the same feature calculated globally for a user.} as well as consider other user behavior features and demographic features~\cite{vigliensoni_automatic_2016}. 
Another research possibility is to evaluate robustness of regression models and study the effect of the size of listening behavior data on the quality of predictions. 

\begin{acks}
This research has been supported by Kakao Corp., 
and partially funded by the European Unions Horizon 2020 research and innovation programme under grant agreement No 688382 (AudioCommons) and 
the  Ministry of Economy and Competitiveness of the Spanish Government (Reference: TIN2015-69935-P). 
\end{acks}

\bibliographystyle{ACM-Reference-Format}
\bibliography{sample-sigconf}

\end{document}